\documentstyle[preprint,aps,epsf]{revtex}
\newif\iftightenlines\tightenlinesfalse
\tightenlines\tightenlinestrue

\def\eslt{\not\!\!{E_T}}
\def\to{\rightarrow}

\def\te{\tilde e}

\def\tu{\tilde u}

\def\tb{\tilde b}

\def\td{\tilde d}

\def\tst{\tilde t}
\def\ttau{\tilde \tau}

\def\tg{\tilde g}
\def\tnu{\tilde\nu}
\def\tell{\tilde\ell}
\def\tq{\tilde q}
\def\tw{\widetilde W}
\def\tz{\widetilde Z}
\newcommand{\met}{\not\!\!\!E_T}
\begin{document}
\draft
\preprint{\vbox{\baselineskip=14pt%
   \rightline{FSU-HEP-010701}
   \rightline{UH-511-994-01}
}}
\title{THE REACH OF THE FERMILAB TEVATRON\\ AND CERN LHC FOR\\
GAUGINO MEDIATED SUSY BREAKING MODELS}
\author{Howard Baer$^1$, Alexander Belyaev$^1$\footnote{On leave of absence from
Nuclear Physics Institute, Moscow State University.},
Tadas Krupovnickas$^1$ and Xerxes Tata$^{2}$}
\address{
$^1$Department of Physics,
Florida State University,
Tallahassee, FL 32306 USA
}
\address{
$^2$Dep't of Physics and Astronomy, University of Hawaii, Honolulu, HI 96822}
\date{\today}
\maketitle
\begin{abstract}

In supersymmetric models with gaugino mediated SUSY breaking (inoMSB),
it is assumed that SUSY breaking on a hidden brane is communicated to
the visible brane via gauge superfields which propagate in the
bulk. This leads to GUT models where the common gaugino mass $m_{1/2}$
is the only soft SUSY breaking term to receive contributions at tree
level. To obtain a viable phenomenology, it is assumed that the gaugino
mass is induced at some scale $M_c$ beyond the GUT scale, and that
additional renormalization group running takes place between $M_c$ and
$M_{GUT}$ as in a SUSY GUT. We assume an $SU(5)$ SUSY GUT above the GUT
scale, and compute the SUSY particle spectrum expected in models with
inoMSB. We use the Monte Carlo program ISAJET to simulate signals within
the inoMSB model, and compute the SUSY reach including cuts and triggers
approriate to Fermilab Tevatron and CERN LHC experiments.  We find no
reach for SUSY by the Tevatron collider in the trilepton channel.
At the CERN LHC, values of $m_{1/2}=1000$ (1160) GeV can be
probed with 10 (100) fb$^{-1}$ of integrated luminosity, corresponding
to a reach in terms of $m_{\tg}$ of 2150 (2500) GeV.
The inoMSB model and mSUGRA can likely only be differentiated at a linear
$e^+e^-$ collider with sufficient energy to produce sleptons and charginos.
\end{abstract}

\medskip



\section{Introduction}

Supersymmetric models with weak scale supersymmetric matter
are very compelling for a variety of reasons, most important of which is that
they solve the gauge hierarchy problem\cite{martin}.
However, it is safe to say that
a compelling model for supersymmetry breaking has yet to emerge.
Supergravity models based on SUSY breaking in a hidden sector\cite{sugra}
can give rise
to weak scale soft SUSY breaking (SSB) terms with SUSY breaking communicated
via gravitational interactions. However, there
exists no compelling mechanism in supergravity to suppress the generation
of non-universal SSB parameters that lead to unacceptably large
flavor violation, sometimes in $CP$-violating processes.
Alternatively, in models with gauge mediated SUSY breaking\cite{gmsb},
universality
of scalars with the same quantum numbers occurs naturally, but at the
expense of the introduction of a messenger sector which acts to
communicate SUSY breaking to the visible sector. Another intriguing
alternative is anomaly-mediated SUSY breaking, an extra dimensional
model wherein SUSY breaking on one brane is communicated to the visible sector
brane via the superconformal anomaly\cite{amsb}.
The form of the scalar masses again allows a solution to the SUSY flavor
(and $CP$) problems.
Alas, the minimal version
of this model leads to tachyonic slepton masses, although the tachyons can be
exorcized in a variety of proposals\cite{amsb2}.

An interesting alternative, also based on extra dimensions, is known as
gaugino-mediated SUSY breaking (inoMSB)\cite{inomsb}.
Like AMSB models, one postulates
the existence of both hidden and visible sector branes, spatially separated
in an extra dimensional world. However, gauge superfields (and perhaps also
Higgs superfields) are allowed to propagate in the bulk. Upon
compactification of the extra dimensions, a tree level
SSB gaugino mass is generated, but all scalar
masses\footnote{We assume that there are no Higgs fields in the
bulk.} and
$A$ (and perhaps $B$) terms are only generated at the loop level, and so are
suppressed, and can be justifiably set to zero at the compactification scale.
The gravitino can be made heavier than the gauginos; it then decouples
and plays no role in our considerations.

Compactification is assumed to occur at or beyond the GUT scale, thus
preserving the successful unification of gauge coupling
constants\cite{ss}.  Thus, at the compactification scale, we expect
\begin{equation}
m_{1/2}\ne0;\ \ \ m_0\sim A_0\simeq 0 ,
\end{equation}
which are the same boundary conditions that arise in no-scale
models\cite{dimitrius}.  If $M_c$ is taken equal to the GUT scale, then
evolution of soft SUSY breaking parameters leads in general to the tau
slepton being the lightest SUSY particle (LSP), a result in conflict
with cosmological considerations, since then the present day universe
would be filled with charged relics, for which there exist stringent
limits.  A way out, proposed by Schmaltz and Skiba\cite{ss}, is that
$M_c>M_{GUT}$, and that above $M_{GUT}$ some four-dimensional GUT gauge
symmetry is valid, such as $SU(5)$ or $SO(10)$. In this case, the
additional beyond-the-GUT-scale RGE running leads to large enough
slepton masses at $M_{GUT}$ that regions of parameter space exist
\cite{ss,bdqt} with a neutralino LSP, in accord with cosmological
constraints.

In the minimal gaugino mediation model, the bilinear SSB term $B\sim 0$.
By minimizing the scalar potential of the MSSM at the weak scale, the
weak scale value of $B$ is related to the parameter $\tan\beta$.  Thus,
in minimal gaugino mediation, the value of $\tan\beta$ is predicted, and
found for instance in Ref.~\cite{ss} to be $\tan\beta\sim 9-22$.  On a
different track, the assumption of a GUT theory above $M_{GUT}$
frequently implies relations amongst the Yukawa couplings of the theory,
especially for the third generation. Thus, in minimal $SU(5)$ we expect
$f_b=f_\tau$ for scales $Q>M_{GUT}$ and in minimal $SO(10)$ we expect
$f_b=f_t=f_\tau$, where the $f_i$ are Yukawa couplings. We adopt the
computer program ISAJET v7.58~\cite{isajet} for calculating RG evolution. Starting
with $\overline{DR}$ fermion masses and gauge couplings at the weak
scale, ISAJET calculates an iterative solution to the relevant set of
RGEs of the MSSM by running between the weak and $GUT$ scales. To test
Yukawa coupling evolution, it is imperative to include SUSY loop
corrections to fermion masses at the weak scale\cite{hall,pierce}. It
has been found that a high degree of $f_b-f_\tau$ Yukawa coupling
unification at $Q=M_{GUT}\sim 2\times 10^{16}$ can occur only for values
of $\tan\beta\sim 30-50$, and $\mu <0$. Similarly, a high degree of
$f_b-f_\tau -f_t$ Yukawa unification only occurs for $\tan\beta\sim 50$
and $\mu<0$\cite{pierce,bdft,deboer,ky}.  We adopt the criteria of
Yukawa coupling unification as being more fundamental than the
generation of tiny GUT scale values of $B$, so that $\tan\beta$
(and the sign of $\mu$) are free parameters, although they are highly
constrained by the requirement of $b-\tau$ unification.

In this paper, we adopt as an example choice a model of gaugino
mediation which reduces to a SUSY $SU(5)$ GUT at the compactification scale.
Our goal is to calculate the spectrum of superpartners at the weak scale,
so that collider scattering events may be generated. We then evaluate
the reach of both the Tevatron $p\bar{p}$ and CERN LHC $pp$
colliders for inoMSB models. In Sec.~\ref{sec:2}
we discuss our results for the spectrum of SUSY particles expected in inoMSB
including $SU(5)$ gauge symmetry below $M_c$.
In Sec.~\ref{sec:3}, we present our
results for the reach of the Fermilab Tevatron for inoMSB models.
In Sec. \ref{sec:4} we present similar results for the CERN LHC.
Finally, we present our conclusions in Sec.~\ref{sec:5}.

\section{Sparticle mass spectrum}\label{sec:2}

If the scale at which SSB terms are generated is substantially higher than
$M_{GUT}$ (but smaller than $M_P$)
then renormalization group (RG) evolution induces a
non-universality at the GUT scale.  The effect can be significant if
large representations are present.  Here, we assume that supersymmetric
$SU(5)$ grand unification is valid at mass scales $Q> M_{GUT}\simeq
2\times 10^{16}$ GeV, extending at most to the reduced Planck scale
$M_P\simeq 2.4\times 10^{18}$ GeV. Below $Q=M_{GUT}$, the $SU(5)$ model
breaks down to the MSSM with the usual $SU(3)_C\times SU(2)_L\times
U(1)_Y$ gauge symmetry. This model is well described in
the work of Polonsky and Pomarol\cite{pp}.

In the $SU(5)$ model, the $\hat{D}^c$ and $\hat{L}$ superfields are
elements of a $\bf{\bar{5}}$ superfield $\hat\phi$, while the $\hat{Q}$,
$\hat{U}^c$ and $\hat{E}^c$ superfields occur in the $\bf 10$
representation $\hat\psi$.  The Higgs sector is comprised of three
super-multiplets: $\hat{\Sigma} ({\bf 24})$ which is responsible for
breaking $SU(5)$, plus $\hat{\cal H}_1(\overline{\bf 5})$ and $\hat{\cal
H}_2({\bf 5})$ which contain the usual Higgs doublet superfields
$\hat{H_d}$ and $\hat{H_u}$ respectively, which occur in the MSSM. The
superpotential is given by
\begin{eqnarray}
\hat{f}=\mu_\Sigma tr\hat{\Sigma}^2 &+&{1\over 6}\lambda'tr\hat{\Sigma}^3 +
\mu_H\hat{\cal H}_1\hat{\cal H}_2 +\lambda\hat{\cal H}_1\hat\Sigma
\hat{\cal H}_2 \nonumber \\
&+&{1\over 4}f_t\epsilon_{ijklm}\hat{\psi}^{ij}\hat{\psi}^{kl}\hat{\cal H}_2^m
+\sqrt{2}f_b\hat{\psi}^{ij}\hat{\phi}_i\hat{\cal H}_{1j} ,
\end{eqnarray}
where a sum over families is understood. $f_t$ and $f_b$ are the top and
bottom quark Yukawa couplings, $\lambda$ and $\lambda'$ are GUT Higgs sector
self couplings, and $\mu_\Sigma$ and $\mu_H$ are superpotential Higgs
mass terms.

Supersymmetry breaking is parametrized by
the SSB terms:
\begin{eqnarray}
{\cal L}_{soft}&=&-m_{{\cal H}_1}^2|{\cal H}_1|^2-
m_{{\cal H}_2}^2|{\cal H}_2|^2-
m_\Sigma^2tr\{ \Sigma^\dagger\Sigma \}-
m_5^2|\phi |^2 - m_{10}^2tr\{\psi^\dagger\psi \}
-{1\over 2}M_5\bar{\lambda}_{\alpha}
\lambda_{\alpha} \nonumber \\
&+&\left[ B_\Sigma\mu_\Sigma tr\Sigma^2 +{1\over 6}A_{\lambda'}\lambda'
tr\Sigma^3 +B_H\mu_H{\cal H}_1{\cal H}_2 +A_\lambda\lambda{\cal H}_1\Sigma
{\cal H}_2 \right. \nonumber \\
&+&\left. {1\over 4}A_tf_t\epsilon_{ijklm}\psi^{ij}\psi^{kl}{\cal H}_2^m
+\sqrt{2}A_bf_b\psi^{ij}\phi_i{\cal H}_{1j}+h.c.\right] ,
\end{eqnarray}
where the fields without the carets denote the appropriate scalar
components.
The various soft masses and gauge and Yukawa couplings evolve with
energy according to the 15 renormalization group equations
given in Ref.~\cite{pp,bdqt}.

To generate the weak scale MSSM mass spectrum, one begins with
the input parameters
\begin{equation}
\alpha_{GUT},\ f_t,\ f_b,\ \lambda ,\ \lambda'
\label{y_su5}
\end{equation}
stipulated at $Q=M_{GUT}$, where $f_b=f_\tau$ is obtained from the
corresponding mSUGRA model.  The first three of these
can be extracted, for instance, from ISASUGRA by
finding points in mSUGRA parameter space with $f_b=f_\tau$.
The
couplings $\lambda (M_{GUT})$ and $\lambda' (M_{GUT})$ are additional
inputs, where $\lambda (M_{GUT})\agt 0.7$~\cite{pdk} to make the triplet
Higgsinos heavy enough to satisfy experimental bounds on the proton
lifetime. The gauge and Yukawa couplings can be evolved via the RGEs to
determine their values at $Q=M_c$.  Assuming universality at $M_c$, we
impose
\begin{eqnarray}
m_{10}&=&m_5=m_{{\cal H}_1}=m_{{\cal H}_2}=m_\Sigma \equiv m_0=0
\nonumber \\
A_t&=&A_b=A_\lambda =A_\lambda'\equiv A_0=0,
\label{MpBoundary}
\end{eqnarray}
as boundary conditions that define our $SU(5)$ inoMSB framework.
We then evolve all the $SU(5)$ soft masses from $M_c$ to $M_{GUT}$.
The MSSM soft breaking masses at $M_{GUT}$ are specified via
\begin{eqnarray}
&m_Q^2=m_U^2=m_E^2\equiv m_{10}^2\,,&
\nonumber\\
&m_D^2=m_L^2\equiv m_5^2\,,&
\label{MGUTboundary}\\
&m_{H_d}^2=m_{{\cal H}_1}^2\,,\quad m_{H_u}^2=m_{{\cal H}_2}^2\,,&
\nonumber
\end{eqnarray}
which can serve as input to ISAJET~\cite{isajet} via the $NUSUGi$ keywords.
Yukawa couplings induce an inter-generation splitting amongst the scalars.
Since there is no splitting amongst the gaugino masses, the
gaugino masses may be taken to be $M_1=M_2=M_3\equiv m_{1/2}$ where
$m_{1/2}$ is stipulated most conveniently at the $GUT$ scale.

In Fig. \ref{mgmrun}, we show the evolution of the various SSB
parameters of the MSSM, starting with the inoMSB boundary
conditions. Here, the unified gaugino mass is taken to be 400~GeV at
$Q=M_{GUT}$. The compactification scale is taken to be
$M_c=10^{18}$~GeV, with $\tan\beta =35$, $\mu<0$, $\lambda =1.0$ and
$\lambda'=0.1$\footnote{Varying the parameters $\lambda$ and $\lambda'$
typically induces small changes only in {\it third} generation scalar
masses, so that other sparticle masses should not be very sensitive to
variations in these parameters.}.  We see that RG evolution results in
GUT scale scalar masses and $A$-parameters that are substantial
fractions of $m_{1/2}$; {\it i.e.} although we have inoMSB boundary
conditions at the scale $M_c$, there are substantial deviations from
these at $M_{GUT}$. While the inter-generation splitting is small, the
splittings between the {\bf 5} and the {\bf 10} dimensional matter
multiplets, as well as between these and the Higgs multiplets is
substantial.

In Fig. \ref{mass}, we show values of various sparticle and Higgs masses,
plus the $\mu$ parameter, as a function of $m_{1/2}$ for
$\tan\beta =35$ and $\mu <0$. In this plot, we adopt
the values of $f_t=0.489$, $f_b=f_\tau= 0.246$, $g=0.703$ at the
scale $M_{GUT}=1.52\times 10^{16}$ GeV. The output values of
$m_{10}$ and $m_5$ for first and third generations, and $m_{H_u}$ and
$m_{H_d}$ serve as GUT scale inputs for ISAJET to generate the
weak scale sparticle masses shown in the figure. We cut off the curves
at $m_{1/2}=285$ GeV below which $m_{\ttau_1}$ becomes less
than $m_{\tz_1}$. Note that the lower limit on $m_{1/2}$ implies
that $m_{\tw_1}\agt 200$ GeV. The following pattern of
sparticle masses occurs:
\begin{equation}
m_{\tz_1}<m_{\tell_R}<m_{\tw_1}<m_{\tell_L}<m_A<m_{\tst_1}<m_{\tb_1}<m_{\tq}<m_{\tg} .
\end{equation}
The large value of $|\mu |$ that occurs means that the $\tz_1$ is mainly
bino-like, and a good candidate for cold dark matter\cite{ss}.

Specific sparticle masses for an $m_{1/2}=400$ GeV case study are shown
in Table~\ref{tab:cases}, along with an mSUGRA model with a universal
GUT scale mass squared that is a weighted average of the corresponding
inoMSB values.  Many aspects of the spectra shown are similar. However,
it is noteworthy that the splitting of the ${\bf 10}$ and ${\bf 5}$
dimensional representations in inoMSB lead to increased right slepton
and decreased left-slepton masses relative to the mSUGRA case. Such a
splitting may be measureable at linear $e^+e^-$ colliders; we discuss
this further in Sec. \ref{sec:5}.

\section{Reach of the Tevatron collider}\label{sec:3}

From the spectra shown in Fig. \ref{mass}, we see that first and second
generation squarks and gluinos
have masses of at least 600 GeV, and hence are inaccessible \cite{wgr}
to Tevatron searches due to low production cross sections.
Sleptons~\cite{slepton} and third generation squarks~\cite{sender} are
also too heavy to be searched for at the Tevatron.
However,
charginos and neutralinos may be light enough that $\tw_1\tw_1$ and
$\tw_1\tz_2$ production serve as the main SUSY production
mechanisms at the Tevatron.

Since $m_{\tw_1}>m_{\ttau_1}$, while $m_{\tw_1}<m_{\tell_L}$, the
chargino dominantly decays via $\tw_1\to \ttau_1\nu_\tau$.
The neutralino $\tz_2$ is mainly wino-like, and so has only a small
coupling to $\tell_R$. Thus, even though the decay mode
$\tz_2\to \tell_R\ell$ is open, the decay mode $\tz_2\to\ttau_1\tau$
is dominant. Thus, we expect signals rich in tau leptons.

The two most promising avenues to explore for Tevatron reach
consist of a clean trilepton search\cite{trilep}, where the focus is on soft
trileptons originating from tau decays, or for trilepton
events where in fact one or more of the identified leptons is
a hadronic tau\cite{ltanb}.

To estimate the Tevatron reach for inoMSB models with soft trileptons,
we adopt the cuts SC2 advocated in the last of
Refs. \cite{bk,mp,new3l}. These cuts have been optimized to maintain
signal while rejecting backgrounds coming from $WZ$ production,
$t\bar{t}$ production and $W^*\gamma^*$ and $W^*Z^*$ production, where
the starred entries correspond to off-shell processes. The cuts include
requiring three isolated\footnote{Leptons with $p_T \ge 5$~GeV
are defined to be isolated if
the hadronic $E_T$ in a $\Delta R$ =0.4 cone about the lepton is
smaller than 2 GeV.} leptons (either $e$s or $\mu$s) with
$p_T(\ell_1 ,\ell_2,\ell_3) > 11,\ 7,\ 5$ GeV respectively, and with
$|\eta(\ell )|<2$, but including at least one lepton with $p_T>11$ GeV
within $|\eta|<1$. In addition, a missing energy cut $\eslt >25$ GeV is
required. Furthermore, a $Z$ veto $m(\ell\bar{\ell})<81$ GeV and a
virtual photon veto $m(\ell\bar{\ell})>20$ GeV is required for opposite
sign/same flavor dilepton pairs. A transverse mass veto $60$ GeV
$<M_T(\ell, \eslt )<85$ GeV is also required to reject on and off shell
backgrounds including $W$ bosons.  The background estimate is then 1.05
fb.

In Fig. \ref{tev3l}, we show the isolated trilepton cross section
after cuts SC2, along with the signal levels needed to achieve
a $5\sigma$ signal at 2 fb$^{-1}$ of integrated luminosity, and a
$5\sigma$ or $3\sigma$ signal at 25 fb$^{-1}$. We see that the
trilepton signal level corresponds to $\alt 1$~trilepton event even
with an integrated luminosity of 25~fb$^{-1}$ and so appears to be
undetectable for the entire range of $m_{1/2}$.

The other possible signal channel is to look for trilepton
events where one or more of the leptons is, in fact, a
tau identified via its hadronic decay.
These can be separated into $\ell\bar{\ell}\tau$ (opposite-sign),
$\ell\ell\tau$ (same-sign), $\ell\tau\tau$ and $\tau\tau\tau$
channels. Signals including tau leptons were first examined in the context
of large $\tan\beta$ in Ref. \cite{ltanb}, and refined background
estimates were presented in Ref. \cite{lykken}.
We adopt the cuts and backgrounds presented in Ref. \cite{lykken}
for our analysis.
Following Ref. \cite{lykken}, we define tau jets to be
hadronic jets with $|\eta|<1.5$, net
charge $\pm 1$, one or three tracks in a $10^\circ$
cone with no additional tracks in a $30^\circ$
cone, $E_T>5$ GeV, $p_T>5$ GeV, plus an electron rejection cut.
The cuts that we implement depend on the event topology, and include:
two isolated ($E_T(cone)<2$ GeV) leptons with $p_T>8$ GeV and $p_T>5$ GeV, and
one identified tau jet with $p_T(\tau )>15$ GeV for  $\ell\bar{\ell}\tau$
and $\ell\ell\tau$
signatures;
two tau jets with $p_T>15$ GeV and  $p_T>10$ GeV
and one isolated lepton with $p_T>7$ GeV for
$\ell\tau\tau$ signature;
three tau jets with
$p_T>15,10$ and 8 GeV, respectively for $\tau\tau\tau$ signature.
For the $\ell\bar{\ell}\tau$ topology, following Ref. \cite{lykken},
we impose additional cuts for same flavor,
opposite sign leptons: $|m(\ell\bar{\ell})-M_Z|>10$ GeV
and $m(\ell\bar{\ell})>11$.
To maximize the signal statistics we
chose set``A)" from the paper~\cite{lykken}: $\met>20$ GeV and no jet veto
requirement.

We consider a signal to be observable if {\it i})~the signal to
background ratio, $S/B \ge 0.2$, {\it ii})~the signal has a minimum of
five events, and {\it iii})~the signal satifies a statistical criterion
$S \ge 5\sqrt{B}$.
Our results for signals including tau leptons are shown in Fig. \ref{tevtau}.
We use the background estimates from Ref.~\cite{lykken} to determine the
minimum signal level for observability. These background cross sections
for $\ell\bar{\ell}\tau$,
$\ell\ell\tau$, $\ell\tau\tau$ and $\tau\tau\tau$ topologies are
10.7~fb, 0.85~fb, 60.4~fb and 24.7~fb, respectively.
We clearly see  that signal is, once
again, too low to be detectable at luminosity upgrades of the Tevatron.
We conclude that in this framework direct detection of sparticles will
not be possible at the Tevatron.


\section{Reach of the CERN LHC}\label{sec:4}

At the CERN LHC, gluino and squark pair production reactions will
be the dominant SUSY production reactions. Gluino and squark
production will be followed by cascade decays \cite{bbkt}, in which a variety
of jets, isolated leptons and missing energy will be produced.
A variety of signals emerge, and can be classified by the number of
isolated leptons present. The signal
channels include {\it i.})~no isolated leptons plus jets plus $\eslt$
($\eslt$),
{\it ii.})~single isolated lepton plus jets plus $\eslt$ ($1\ell$),
{\it iii.})~two opposite sign isolated leptons plus jets plus $\eslt$
(OS), {\it iv.})~two same sign isolated leptons plus jets plus $\eslt$
(SS) and {\it v.})~three isolated leptons plus jets plus $\eslt$
($3\ell$).

The reach of the CERN LHC for SUSY has been estimated for the
mSUGRA model in Ref. \cite{bcpt1,cms} at low $\tan\beta$ and in
Ref. \cite{lhcltanb} at large $\tan\beta$.
We adopt the cuts and background levels presented in
Ref. \cite{bcpt1} for our analysis of the signal channels listed above.
Hadronic clusters with
$E_T>100$ GeV and $|\eta ({\rm jet})|<3$ within a cone of size $R=\sqrt{\Delta\eta^2
+\Delta\phi^2} =0.7$ are classified as jets.
Muons and electrons are classified as isolated if they have $p_T>10$ GeV,
$|\eta (\ell )|<2.5$, and the visible activity within a cone of $R
=0.3$ about the lepton direction is less than $E_T({\rm cone})=5$ GeV.

Following Ref. \cite{bcpt1}, we required that the jet multiplicity,
$n_{\rm jet} \geq 2$, transverse sphericity $S_T > 0.2$, $E_T(j_1)$, and
further, that $E_T(j_2) \ > \ E_T^c$ and $\eslt > E_T^c$, where the cut
parameter $E_T^c$ is chosen to roughly optimize the signal from gluino
and squark production.  For the leptons we require $p_T(\ell) > 20$~GeV
($\ell=e$ or $\mu$) and $M_T(\ell,\eslt) > 100$~GeV for the $1\ell$
signal. For the $OS$, $SS$ and $3\ell$ channels, we require that the two
hardest leptons have $p_T \ge 20$~GeV.
We have also applied the cut on the
transverse plane angle $\Delta\phi (\vec{\eslt},j_c )$
between $\vec{\eslt}$ and closest jet: $30^\circ<\Delta\phi <90^\circ$,
in the case of the $\eslt$ channel, $i)$.

Our results for the $\eslt$ signal channel are shown in Fig. \ref{lhc1}
for choices of the cut parameter $E_T^c =100$, 300 and, for the $\eslt$
and $1\ell$ channels, also 500~GeV.  The error bars denote the
statistical uncertainty in our Monte Carlo calculation.  The solid
(dashed) horizontal mark on each curve denotes the minimum cross section
needed for discovery,
incorporating the three criteria listed in the
last section, for an integrated luminosity of 10 (100)~fb$^{-1}$.
For those values of $E_T^c$ where the reach is limited by the $S/B \ge
0.2$ requirement, increasing the integrated luminosity does not improve
the reach, and we have no dashed horizontal line.
Although the signal is largest for the softer cuts, larger
$E_T^c$ values (corresponding to harder cuts) are more effective in
selecting signal events over background for very heavy squarks and gluinos.
This is why the reach is maximized for the largest $E_T^c$ value for
which the signal still leaves an observable number of events. Thus, in
the $\eslt$ channel, the $5\sigma$ reach is found to be 925 (1100)~GeV
in the parameter $m_{1/2}$ for 10 (100)~fb$^{-1}$.  This corresponds to
a reach in $m_{\tg}$ of 2000 (2400)~GeV, respectively.

The corresponding situation for the $1\ell$ channel is shown in
Fig.~\ref{lhc2}. Once again, the largest reach is obtained for
$E_T^c=500$ GeV.  We see that $m_{1/2}=1000$ (1160) GeV should be
accessible for 10 (100) fb$^{-1}$ of integrated luminosity,
corresponding to a reach in $m_{\tg}$ of 2150 (2500) GeV.

For channels with $\ge 2$ leptons, we have conservatively restricted our
analyis to $E_T^c$ smaller than 300~GeV, because for larger values of
this cut parameter, the estimates of the SM backgrounds may have
considerable statistical fluctuations.  The results for the opposite
sign ($OS$) dilepton channel is shown in Fig. \ref{lhc3}, where the reach with
$E_T^c=300$ GeV is found to be $m_{1/2}=750$ (900)~GeV for 10
(100)~fb$^{-1}$ of integrated luminosity, corresponding to a reach in
$m_{\tg}$ of $\sim$1650 (1950) GeV.
The expectation for the same sign ($SS$) dilepton channel is shown in
Fig. \ref{lhc4}, where the reach with $E_T^c=300$ GeV is found to be
$m_{1/2}=800$ (925)~GeV for 10 (100)~fb$^{-1}$ of integrated
luminosity, corresponding to a reach in $m_{\tg}$ of 1700 (2000) GeV.
Finally, for the $3\ell$ channel shown in Fig. \ref{lhc5}, the reach with
$E_T^c=300$ GeV is found to be essentially the same as that for the $SS$
channel.

Thus, for inoMSB, we expect a robust signal for SUSY in a variety
of channels, with the $1\ell$ channel offering the best
ultimate reach for SUSY, corresponding to $m_{\tg}$ up to 2.1-2.5~TeV,
for an integrated luminosity of 10-100~fb$^{-1}$.

\section{Conclusions}\label{sec:5}

In this paper we have addressed the question of discovery reach for
SUSY breaking models with gaugino mediated SUSY breaking.
These models give rise to ``no-scale'' boundary conditions for
SSB parameters. The boundary conditions are assumed valid at a scale
$M_c$ beyond the GUT scale, but somewhat below the Planck scale. A four
dimensional SUSY GUT model is assumed valid at these high scales, and
for definiteness, we chose a model based on $SU(5)$ gauge
symmetry. Simple models based on $SO(10)$ gauge symmetry are
more difficult to accommodate, since they must obey the
more stringent condition of $t-b-\tau$ Yukawa coupling.
Such Yukawa unified solutions are difficult to reconcile with
the constraint of radiative EWSBW and no-scale type boundary
conditions.

We found that the Fermilab Tevatron has {\it no reach} for sparticles in
the inoMSB model. This occurs for several reasons. First, gluinos and
squarks are beyond the reach of the Tevatron. Second, charginos and
neutralinos dominantly decay to third generation leptons and sleptons,
and taus are more difficult to detect that $e$s and $\mu$s.  Moreover,
the lower limit on parameter space implies $m_{\tw_1}\agt 200$ GeV, so
there is not a lot of sparticle production cross section to begin
with. Finally, since $m_{\ttau_1}\simeq m_{\tz_1}$ at the lower values
of allowed $m_{1/2}$, the stau decays give rise to very soft visible
decay products, reducing greatly the efficiency to detect signals
including hadronic taus.

On the other hand, the CERN LHC has a substantial reach for inoMSB
models. In this case, we expect gluino and squark pair production to
dominate, so that a variety of cascade decay signals will be present if
$m_{1/2}$ is not too large.  We find the greatest reach via the $\eslt$
and $1\ell$ channels, where there should be observable signals for
gluinos as heavy as 2150 (2500)~GeV for 10 (100) fb$^{-1}$ of integrated
luminosity.  If gluinos are lighter than $\sim 1700$ (2000)~GeV, there
should be confirmatory signals also in the OS, SS and $3\ell$ channels.
These reach values are comparable to expectations within the mSUGRA
framework, which is not surprising in that the sparticle mass spectra
for the two models are not that different. Although we have performed
our analysis assuming that there are no Higgs fields in the bulk, we
expect that our conclusions about the LHC reach will be qualitatively
unchanged even if this is not the case. The reason is that gaugino and
scalar masses for the first two generations are insensitive to our
boundary condition $m_{H_u}=m_{H_d}=0$. We also expect that our
estimates of the SUSY reach of the LHC are insensitive to the couplings
in the $GUT$ Higgs sector.

The question then arises whether the inoMSB and mSUGRA models can
be differentiated by collider experiments.
The main spectral difference between the two models arises due to
the non-universal GUT scale scalar masses arising in the inoMSB
model. Most noticeably, the splitting between the ${\bf 10}$ and ${\bf 5}$ of
$SU(5)$ gives rise to heavier right-sleptons and
lighter left-sleptons in the inoMSB model compared to mSUGRA
with a similar overall spectrum (see Table~\ref{tab:cases}).
Differentiation of the two models is a very difficult
task to accomplish at the CERN LHC.

However, a method has recently been proposed to differentiate models
with GUT scale scalar mass non-universality from the mSUGRA framework at
$e^+e^-$ linear colliders\cite{hess}.
These authors have proposed that the measurable quantity
\begin{equation}
\Delta =
m_{\te_R}^2-m_{\te_L}^2+{{m_{\tw_1}^2}\over {2\alpha_2^2(m_{\tw_1})}}
\left[{3\over 11}(\alpha_1^2(m_{\te})-\alpha_1^2(M_{GUT}))
-3(\alpha_2^2(m_{\te}) -\alpha_2^2(M_{GUT}))\right],
\label{delta}
\end{equation}
could be used to differentiate the two classes of models.  This quantity
$\Delta$ is expected~\footnote{Note that the theoretical expectation for
$\Delta$ is slightly different than shown in Ref.~\cite{hess} because
the quantity $S_{GUT}$ (defined therein) is now just $m_{{\cal H}_2}^2
-m_{{\cal H}_1}^2$.} to be small, within the range $-4000$ GeV$^2<\Delta
<2000$ GeV$^2$ for mSUGRA, while it is expected to be much larger in the
inoMSB framework. For instance, for the case study in Table
\ref{tab:cases}, $\Delta\sim 15,000$ GeV$^2$. The difference is
sufficiently large so that the two models should be distinguishable via
precision measurements at a linear $e^+e^-$ collider where chargino and
selectron masses can be determined to about 1-2\%.

In conclusion, if nature has chosen to be described by inoMSB
model with an $SU(5)$ gauge symmetry above the GUT scale, then
we may expect a SUSY Higgs boson discovery at the luminosity upgrade of
the Tevatron, but no sign
of sparticles. Conversely, we would expect a SUSY discovery at the CERN LHC
(unless sparticle masses are so heavy they are in the fine-tuned region
of SUSY parameter space,
with $m_{\tg}>2500$ GeV). However, the underlying model will not be
revealed until a sufficient data set has been accumulated at a linear
$e^+e^-$ collider, where precision measurements of sparticle masses
would point to an inoMSB model with a $SU(5)$ GUT symmetry.

%
\acknowledgments
This research was supported in part by the U.~S. Department of Energy
under contract numbers DE-FG02-97ER41022 and DE-FG-03-94ER40833.
%
%

\newpage
%
%

\iftightenlines\else\newpage\fi
\iftightenlines\global\firstfigfalse\fi
\def\dofig#1#2{\epsfxsize=#1\centerline{\epsfbox{#2}}}

\begin{table}
\begin{center}
\caption{GUT scale SSB parameters and weak scale sparticle
masses and parameters (GeV) for
mSUGRA and inoMSB case studies with $m_{1/2}=400$ GeV, $\tan\beta =35$
and $\mu <0$.}
\bigskip
\begin{tabular}{lcc}
\hline
parameter & mSUGRA & inoMSB  \\
\hline

$m_{10}(1)$ & 205.2 & 233.3 \\
$m_{10}(3)$ & 205.2 & 226.8 \\
$m_{5}(1)$ &  205.2 & 190.5 \\
$m_{5}(3)$ &  205.2 & 188.5 \\
$m_{H_d}$ & 205.2 & 134.9 \\
$m_{H_u}$ & 205.2 & 128.0 \\
$A_t$ & -148.4  & -157.9 \\
$A_b$ & -148.4  & -139.0 \\
$f_t(M_{GUT})$ & 0.497 & 0.489 \\
$f_b(M_{GUT})$ & 0.287 & 0.246 \\
$m_{\tg}$ & 916.6 & 919.5 \\
$m_{\tu_L}$ & 836.5 & 843.8 \\
$m_{\td_R}$ & 805.5 & 801.8 \\
$m_{\tst_1}$& 622.0 & 629.4 \\
$m_{\tb_1}$ & 691.2 & 689.6 \\
$m_{\tell_L}$ & 340.6 & 331.8 \\
$m_{\tell_R}$ & 256.3 & 279.8 \\
$m_{\tnu_{e}}$ & 331.1 & 322.1 \\
$m_{\ttau_1}$ & 193.0 & 210.5  \\
$m_{\tnu_{\tau}}$ & 318.9 & 310.0 \\
$m_{\tw_1}$ & 303.5 & 304.9 \\
$m_{\tz_2}$ & 303.3 & 304.8 \\
$m_{\tz_1}$ & 162.4 & 162.5 \\
$m_h      $ & 117.7 & 117.7 \\
$m_A      $ & 376.1 & 379.8 \\
$m_{H^+}  $ & 386.8 & 390.3 \\
$\mu      $ & -515.1 & -539.3 \\
\hline
\label{tab:cases}
\end{tabular}
\end{center}
\end{table}

\newpage
%

%
\begin{figure}
\dofig{4in}{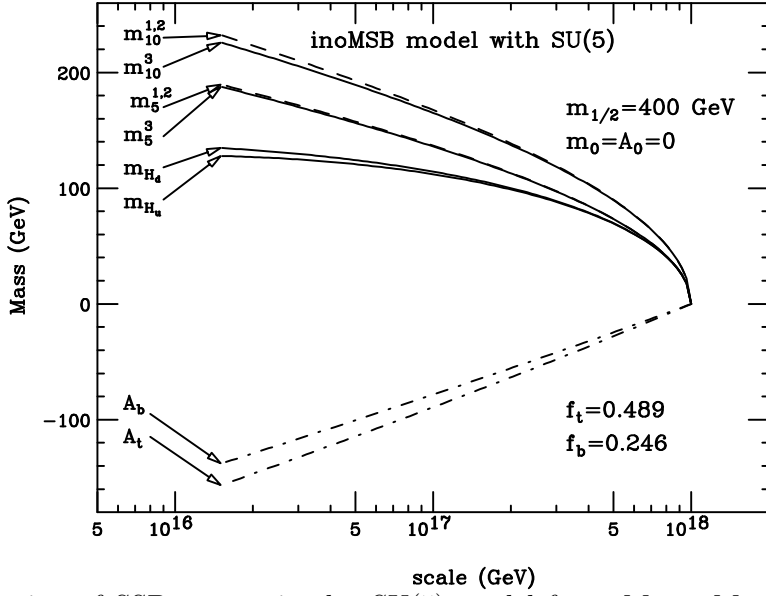}
\caption[]{
Evolution of SSB masses in the $SU(5)$ model from $M_c$ to
$M_{GUT}$, for $\tan\beta =35$, $\mu <0$, $\lambda =1.0$ and
$\lambda'=0.1$, for $m_{1/2} =400$ GeV.}
\label{mgmrun}
\end{figure}
\begin{figure}
\dofig{4in}{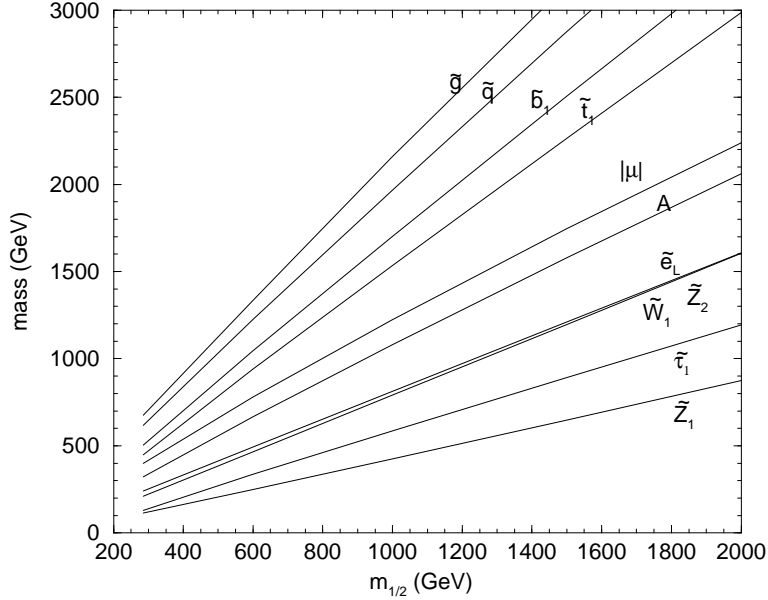}
\caption[]{
Mass values of various SUSY particles and $\mu$ parameter in the
$SU(5)$ inoMSB model with $\tan\beta =35$ and $\mu <0$ versus
the GUT scale common gaugino mass $m_{1/2}$. The lighter chargino and
$\tz_2$ are essentially degenerate, and $\te_L$ is slightly heavier. }
\label{mass}
\end{figure}
\begin{figure}
\dofig{4in}{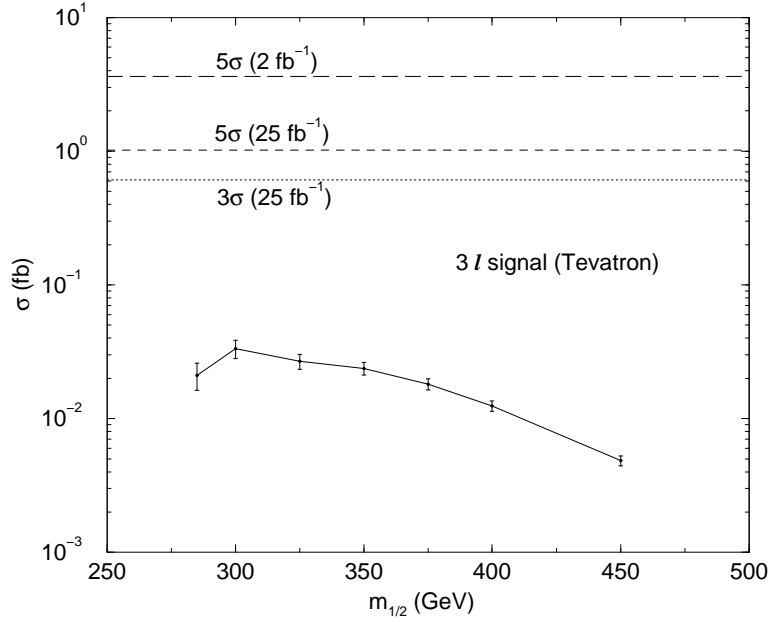}
\caption[]{
Cross section after cuts SC2 of Ref.~\cite{new3l} for trilepton events
at the Fermilab Tevatron. The horizontal lines denote the minimum cross
section for the signal to be observable.}
\label{tev3l}
\end{figure}
\begin{figure}
\dofig{4in}{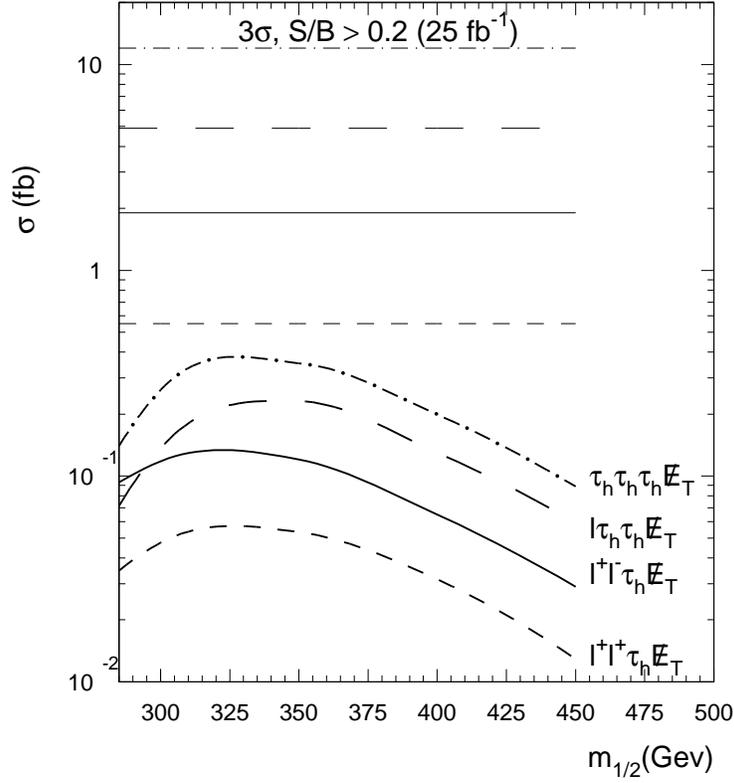}
\caption[]{
Cross section after cuts of Ref.~\cite{lykken} for trilepton events including
identified hadronically decaying tau leptons at the Fermilab
Tevatron. The horizontal lines denote the minimum cross section for the
signal to be observable.}
\label{tevtau}
\end{figure}
\begin{figure}
\dofig{4in}{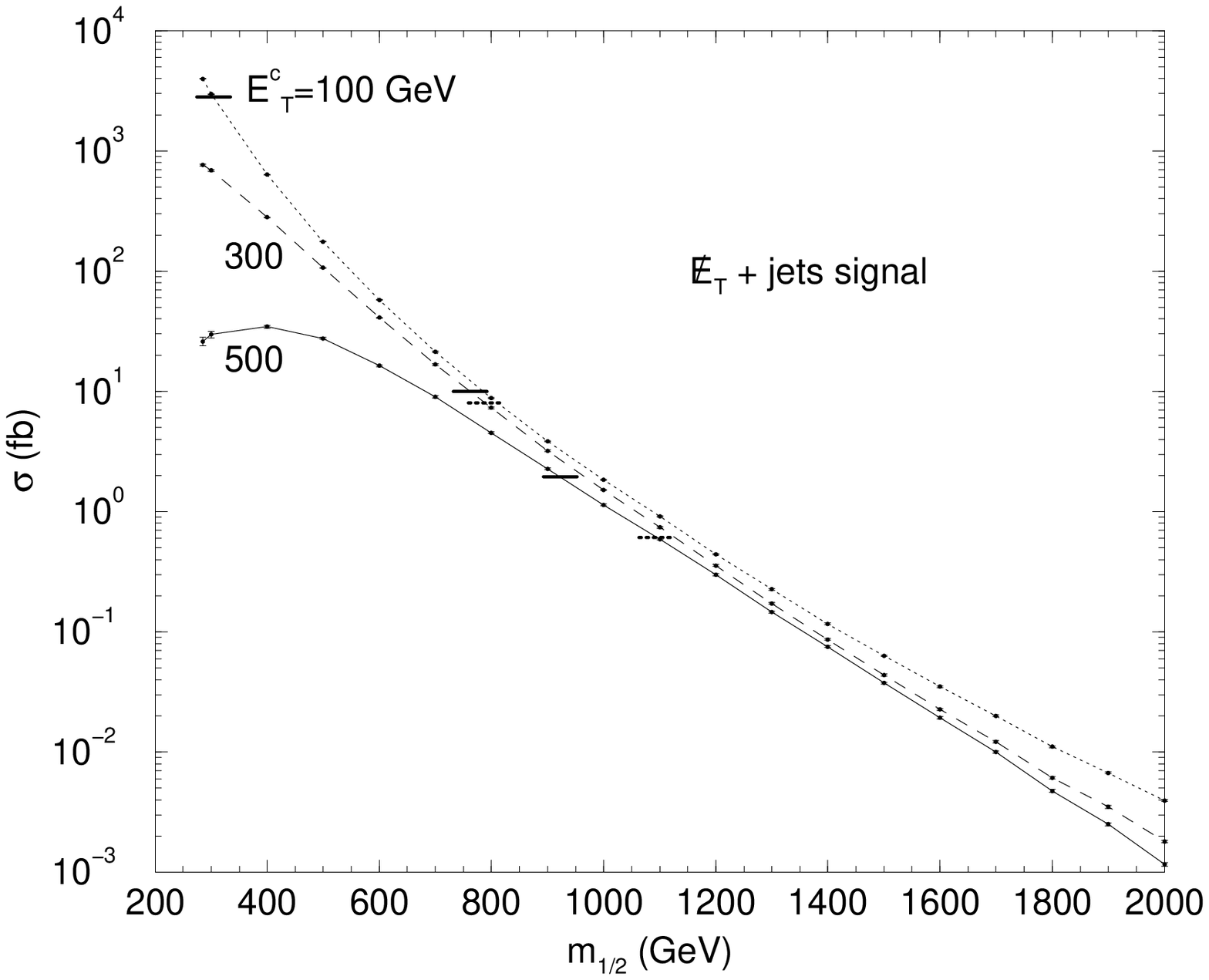}
\caption[]{ Cross section after cuts of Ref.~\cite{bcpt1}
for $\eslt +jets$ events at the
CERN LHC for $E_T^c$ values of 100, 300 and 500 GeV. For each $E_T^c$
value, the reach is
given by the horizontal solid (dashed) line for 10
(100)~fb$^{-1}$ of integrated luminosity.}
\label{lhc1}
\end{figure}
\begin{figure}
\dofig{4in}{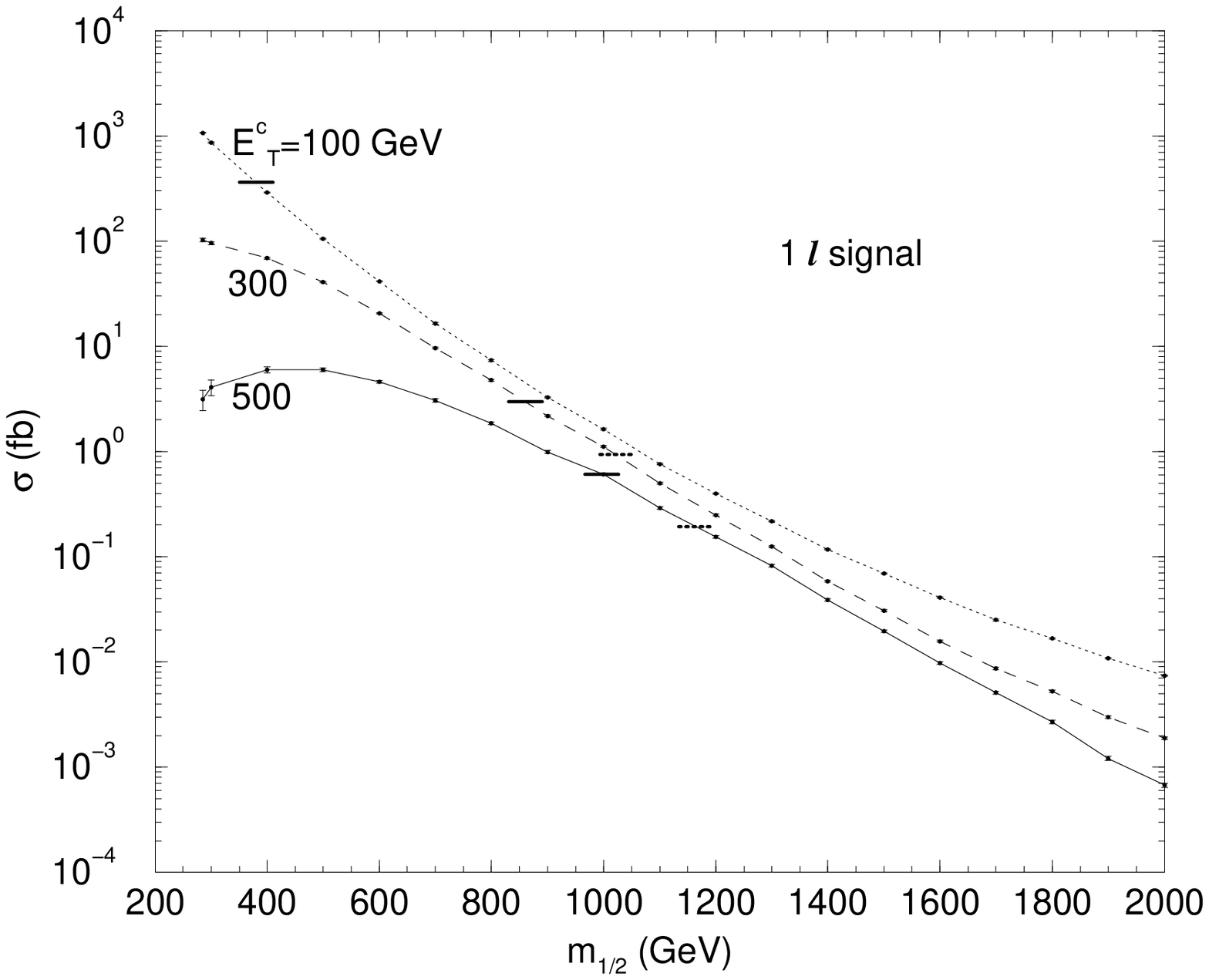}
\caption[]{ Cross section after cuts of Ref.~\cite{bcpt1} for
$1\ell+\eslt +jets$ events at the CERN LHC for $E_T^c$ values of 100,
300 and 500 GeV. For each $E_T^c$ value, the reach is given by the
horizontal solid (dashed) line for 10 (100)~fb$^{-1}$ of integrated
luminosity.}
\label{lhc2}
\end{figure}
\begin{figure}
\dofig{4in}{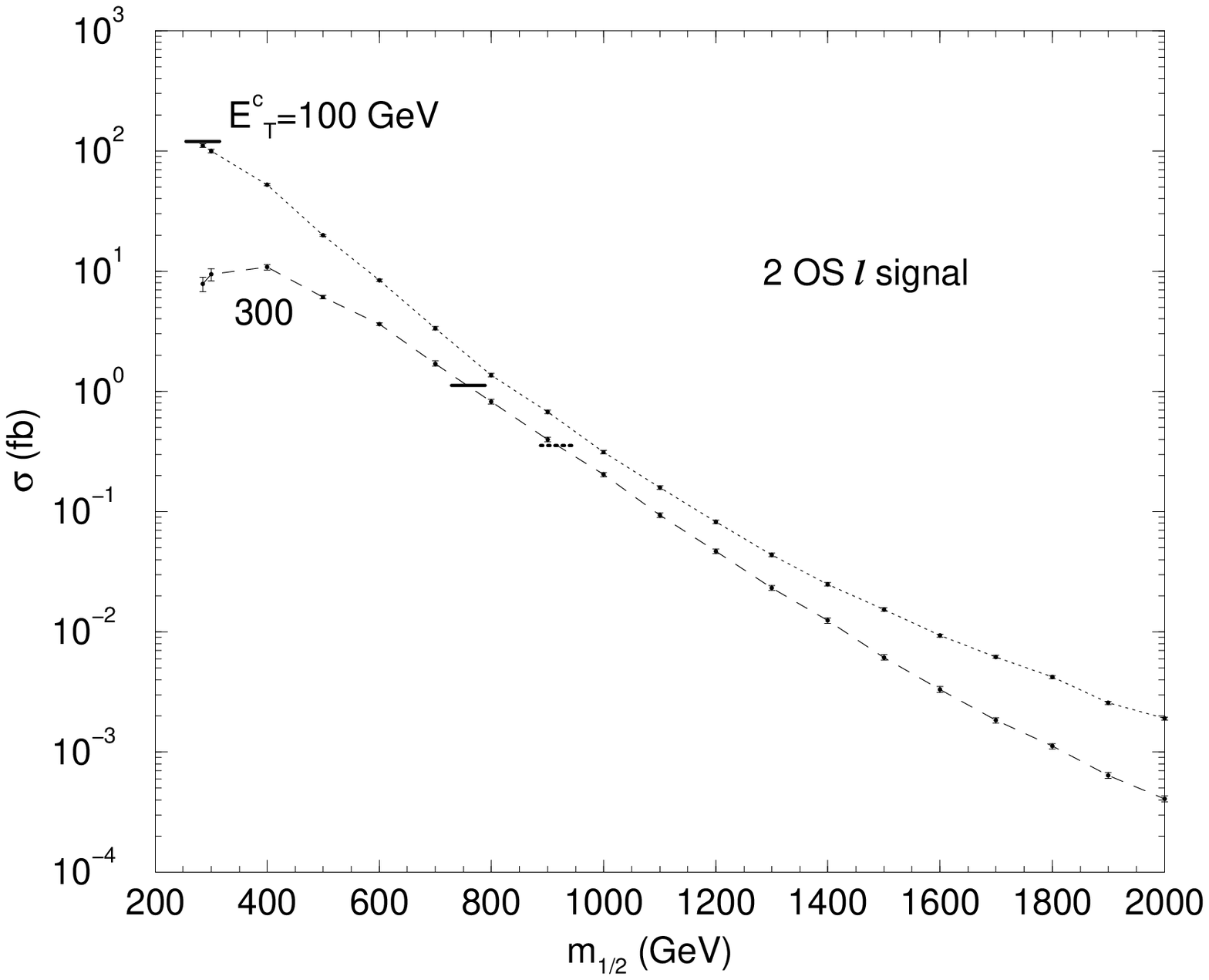}
\caption[]{ Cross section after cuts of Ref.~\cite{bcpt1} for OS
dilepton$+\eslt +jets$ events at the CERN LHC for $E_T^c$ values of 100
and 300 GeV.  For each $E_T^c$ value, the reach is given by the
horizontal solid (dashed) line for 10 (100)~fb$^{-1}$ of integrated
luminosity.}
\label{lhc3}
\end{figure}
\begin{figure}
\dofig{4in}{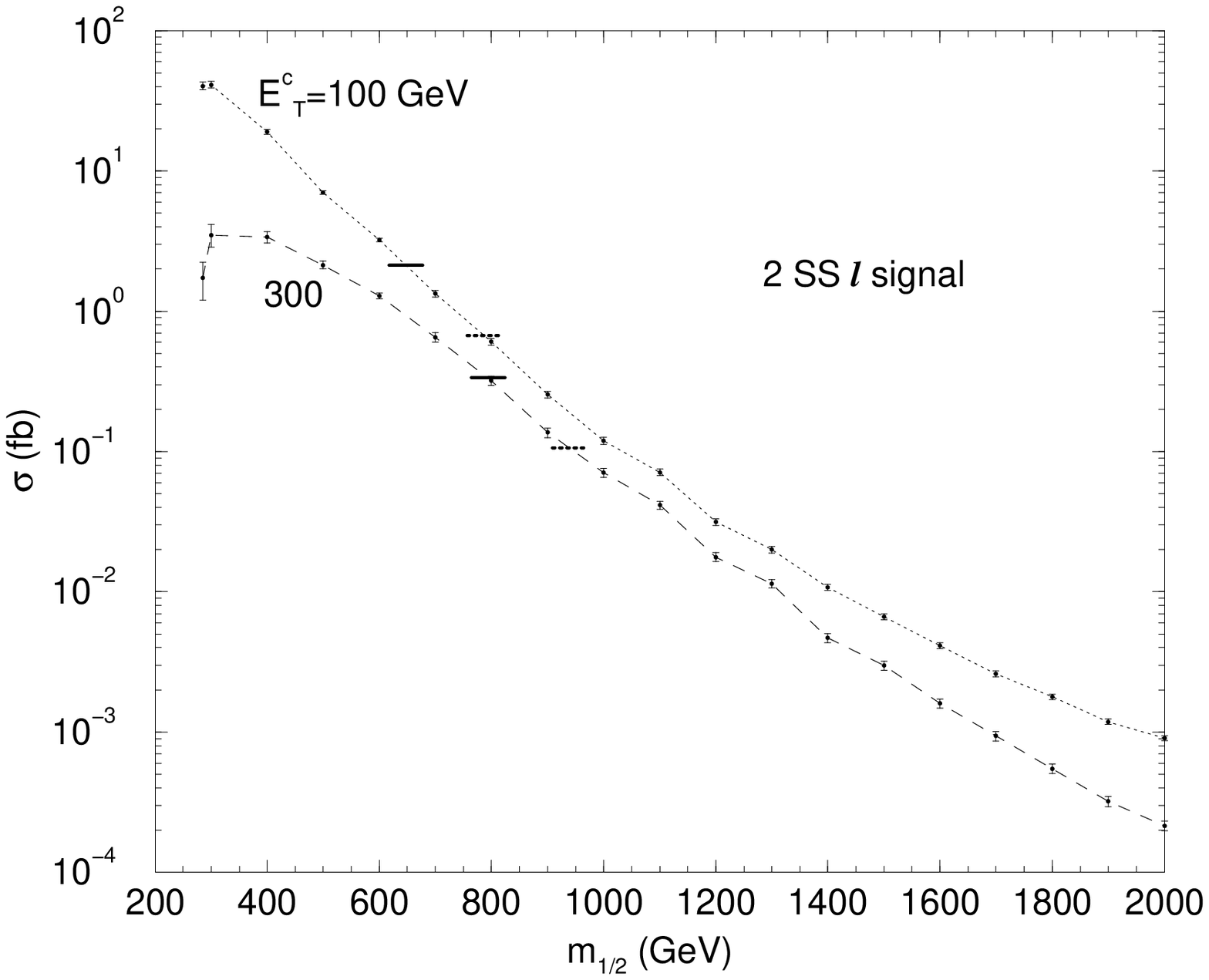}
\caption[]{ Cross section after cuts of Ref.~\cite{bcpt1} for SS
dilepton$+\eslt +jets$ events at the CERN LHC for $E_T^c$ values of 100
and 300 GeV. For each $E_T^c$ value, the reach is given by the
horizontal solid (dashed) line, for 10 (100)~fb$^{-1}$ of integrated
luminosity.}
\label{lhc4}
\end{figure}
\begin{figure}
\dofig{4in}{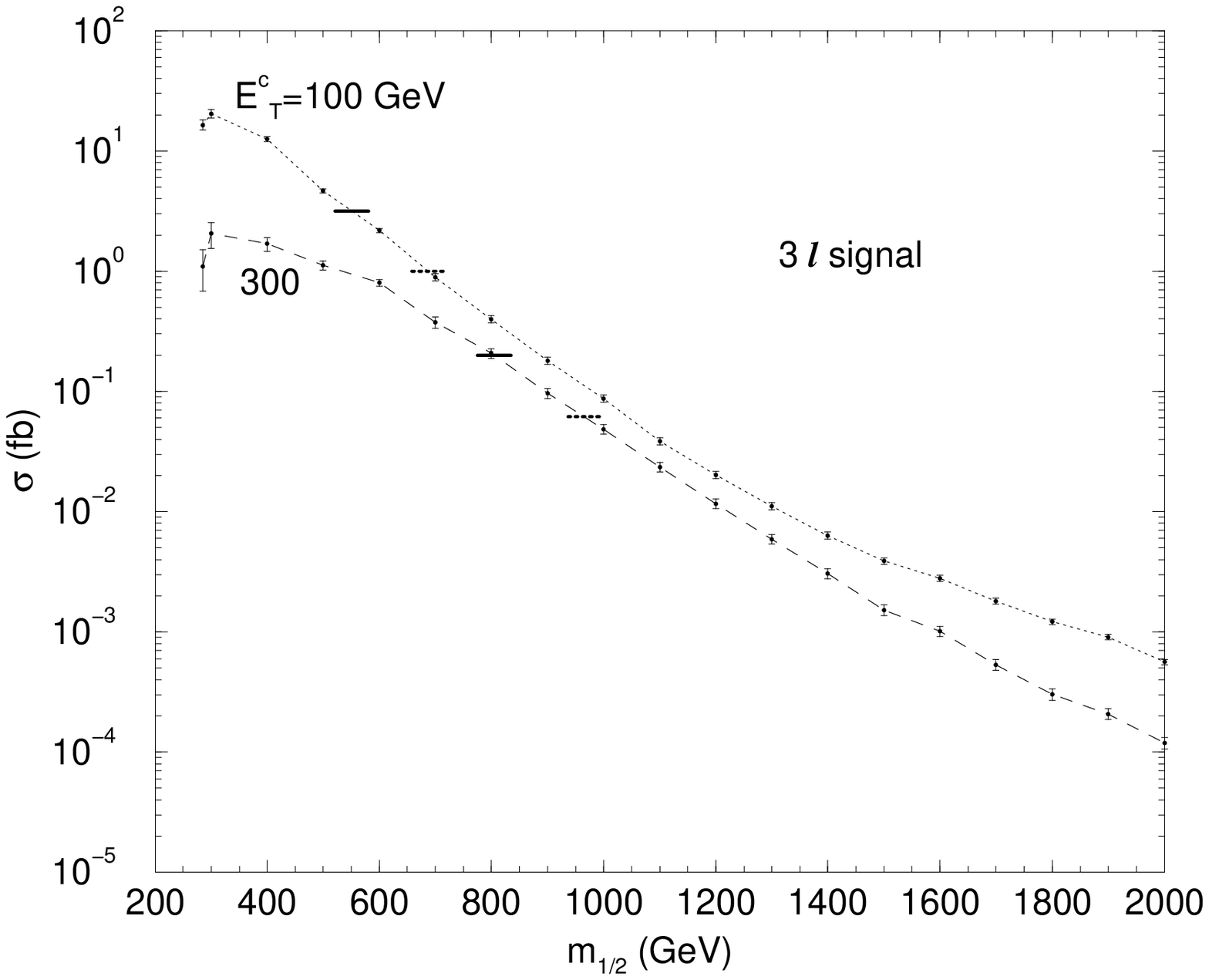}
\caption[]{ Cross section after cuts of Ref.~\cite{bcpt1} for $3\ell
+\eslt +jets$ events at the CERN LHC for $E_T^c$ values of 100 and 300
GeV. For each $E_T^c$ value, the reach is given by the horizontal solid
(dashed) line for 10 (100)~fb$^{-1}$ of integrated luminosity.}
\label{lhc5}
\end{figure}


\begin{references}
\bibitem{martin}For recent reviews, see {\it e.g.} S. Martin,
in {\it Perspectives on Supersymmetry}, edited by G. Kane (World Scientific),
hep-ph/9709356; M. Drees, hep-ph/9611409 (1996);
J. Bagger, hep-ph/9604232 (1996);
X. Tata, {\it Proc. IX J. Swieca Summer School,}
J. Barata, A. Malbousson and S. Novaes, Eds. hep-ph/9706307;
S. Dawson, {Proc. TASI 97}, J. Bagger, Ed. hep-ph/9712464.

%
\bibitem{sugra} A. Chamseddine, R. Arnowitt and P. Nath,
Phys. Rev. Lett. {\bf 49}, 970 (1982); R. Barbieri, S. Ferrara and C. Savoy,
Phys. Lett. {\bf 119B}, 343 (1982); L. J. Hall, J. Lykken and
S. Weinberg, Phys. Rev. D{\bf 27}, 2359 (1983).
%
\bibitem{gmsb} M. Dine, A. Nelson, Y. Nir and Y. Shirman,
Phys. Rev. D{\bf 53}, 2658 (1996); for a review, see
G. Giudice and R. Rattazzi, Phys. Rept. {\bf 322}, 419 (1999).
%
\bibitem{amsb} L. Randall and R. Sundrum,
Nucl. Phys. {\bf B557}, 79 (1999);
G.~F.~Giudice, M.~A.~Luty, H.~Murayama and R.~Rattazzi,
JHEP {\bf 9812}, 027 (1998).
%
\bibitem{amsb2}  See, for instance,
A. Pomarol and R. Rattazzi, JHEP {\bf 05}, 013 (1999);
E. Katz, Y. Shadmi and Y. Shirman, JHEP {\bf 08}, 015 (1999);
R. Rattazzi, A. Strumia and J. Wells, Nucl. Phys. {\bf B576}, 3 (2000);
I. Jack and D.~R.~T.~Jones, Phys. Lett. {\bf B482}, 167 (2000).
%
\bibitem{inomsb} D.~E.~Kaplan, G.~D.~Kribs and M.~Schmaltz,
Phys.\ Rev.\ D {\bf 62}, 035010 (2000);
Z.~Chacko, M.~A.~Luty, A.~E.~Nelson and E.~Ponton,
JHEP {\bf 0001}, 003 (2000)

%
\bibitem{ss}  M.~Schmaltz and W.~Skiba, Phys. Rev. D{\bf 62}, 095005 (2000)
and  Phys. Rev. D{\bf 62}, 095004 (2000).
%
\bibitem{dimitrius} A. B. Lahanas and D. V. Nanopoulos,
Phys. Rept. {\bf 145}, 1 (1987).
%
\bibitem{bdqt} H. Baer, M. Diaz, P. Quintana and X. Tata,
JHEP {\bf 0004}, 016 (2000).
%
\bibitem{isajet} F. Paige, S. Protopopescu, H. Baer and X. Tata,
hep-ph/0001086 (2000).
%
\bibitem{hall} L. J. Hall, R. Rattazzi and U. Sarid,
Phys. Rev. D{\bf 50}, 7048 (1994);
R. Hempfling, Phys. Rev. D{\bf 49}, 6168 (1994); M. Carena, M. Olechowski,
S. Pokorski and C. Wagner, Nucl. Phys. {\bf B426}, 269 (1994).
%
\bibitem{pierce} D. Pierce, J. Bagger, K. Matchev and R. Zhang,
Nucl. Phys. {\bf B491}, 3 (1997).
%
\bibitem{bdft} H. Baer, M. Diaz, J. Ferrandis and X. Tata,
Phys. Rev. D{\bf 61}, 111701 (2000).
%
\bibitem{deboer}
W.~de Boer, M.~Huber, C.~Sander and D.~I.~Kazakov,
Phys.\ Lett.\ B {\bf 515}, 283 (2001).
%
\bibitem{ky} S. Komine and M. Yamaguchi, hep-ph/0110032 (2001).
%
\bibitem{pp}  N. Polonsky and A. Pomarol,
Phys. Rev. D {\bf 51}, 6532 (1995).
%
\bibitem{pdk} R. Arnowitt and P.~Nath, Phys. Rev. Lett. {\bf 69}, 725
(1992); J.~Hisano, H.~Murayama and T.~Yanagida, Nucl. Phys. B {\bf 402},
46 (1993).
%
\bibitem{wgr} S.~Abel {\it et al.} Report of SUGRA Working Group for
Run~II of the Tevatron, hep-ph/0003154 (2000).
%
\bibitem{slepton} H.~Baer, C.H.~Chen, F. Paige and X.~Tata,
Phys. Rev. {\bf D49}, 3283 (1994).
%
\bibitem{sender} H.~Baer, J.~Sender and X.~Tata, Phys. Rev. {\bf D50},
4517 (1994); J.~Sender,  Ph.D thesis, hep-ph/0010025;
H.~Baer, P.~Mercadante and X.~Tata, Phys. Rev. {\bf D59}, 015010 (1999);
R.~Demina, J.~Lykken, K.~Matchev and A. Nomerotski, Phys. Rev. {\bf
D62}, 035011 (2000).
%
\bibitem{trilep} H. Baer, K. Hagiwara and X. Tata,
Phys. Rev. D{\bf 35}, 1598 (1987); R. Arnowitt and P. Nath,
Mod. Phys. Lett. {\bf A2}, 331 (1987); H. Baer and X. Tata,
Phys. Rev. D{\bf 47}, 2739 (1993).
%
\bibitem{ltanb} H. Baer, C. H. Chen, M. Drees, F. Paige and X. Tata,
Phys. Rev. Lett. {\bf 79}, 986 (1997) and
Phys. Rev. D{\bf 58}, 075008 (1998).
%
\bibitem{bk} V. Barger, C. Kao and T. Li, Phys. Lett. {\bf B433}, 328 (1998);
V. Barger and C. Kao, Phys. Rev. D{\bf 60}, 115015 (1999).
%
\bibitem{mp} K. Matchev and D. Pierce, Phys. Rev. D{\bf 60}, 075004 (1999)
and Phys. Lett. {\bf B467}, 225 (1999).
%
\bibitem{new3l} H. Baer, M. Drees, F. Paige, P. Quintana and X. Tata,
Phys. Rev. D{\bf 61}, 095007 (2000).
%
\bibitem{lykken} J. Lykken and K. Matchev, Phys. Rev. D{\bf 61},
015001 (1999).
%
\bibitem{bbkt} H.~Baer, V.~Barger, D.~Karatas and X.~Tata,
Phys. Rev. {\bf D36}, 96 (1987).
%
\bibitem{bcpt1} H. Baer, C. H. Chen, F. Paige and X. Tata,
Phys. Rev. D{\bf 52}, 2746 (1995) and Phys. Rev. D{\bf 53}, 6241 (1996).
%
\bibitem{cms} S. Abdullin {\it et al.} (CMS Collaboration),
hep-ph/9806366 (1998); S. Abdullin and F. Charles,
Nucl. Phys. {\bf B547}, 60 (1999); ATLAS collaboration, Technical Design
Report, V II, CERN/LHCC/99-15 (1999).
%
\bibitem{lhcltanb} H. Baer, C. H. Chen, M. Drees, F. Paige and X. Tata,
Phys. Rev. D{\bf 59}, 055014 (1999).
%
\bibitem{hess} H. Baer, C. Balazs, S. Hesselbach, J. K. Mizukoshi
and X. Tata, Phys. Rev. D{\bf 63}, 095008 (2001).
%
\end{references}
\end{document}